\documentclass[showpacs,twocolumn,prhrenumbers,amsmath,amssymb]{revtex4}
\usepackage{amsmath,amsfonts,latexsym,amssymb,graphicx,graphics,epsfig,subfigure,color,makeidx,xcolor,epstopdf,multirow}
\usepackage[colorlinks=true,citecolor=blue,linkcolor=red,anchorcolor=green,urlcolor=cyan]{hyperref}
\usepackage[title]{appendix}
\newcommand {\nn}{\nonumber}

\newcommand {\be}{\begin{equation}}
\newcommand {\ee}{\end{equation}}
\newcommand {\beq}{\begin{eqnarray}}
\newcommand {\eeq}{\end{eqnarray}}

\graphicspath{{plots/}}
\makeatletter

\begin{document}
\title{The properties and predictions of quasi-periodic oscillations around a black hole in nonlocal gravity}

\author{Tao-Tao Sui$^{a}$\footnote{suitaotao@aust.edu.cn}}
\author{Chen Long$^{a}$}
\author{Ye zhang$^{a}$}

\affiliation{$^{a}$Center for Fundamental Physics, School of Mechanics and Photoelectric Physics, Anhui University of Science and Technology, Huainan, Anhui 232001,  China.
}

\begin{abstract}
{We investigate the dynamics of massive test particles around a static black hole in nonlocal gravity and examine the corresponding properties of high-frequency quasi-periodic oscillations (HF QPOs), constraining the nonlocal parameter to {$\alpha/M \leq 0.452$}. We show that the nonlocal parameter $\alpha$ enhances the effective potential $V_{\text{eff}}$ and leads to a systematic reduction in the energy $\mathcal{E}$ and angular momentum $\mathcal{L}$ of circular orbits. Consequently, the innermost stable circular orbit (ISCO) radius, along with the associated energy and angular momentum, decreases monotonically with $\alpha$, while the radiative efficiency increases, reaching a maximum of approximately $8.9\%$. We further analyze the fundamental orbital frequencies of test particles and find that, due to the spherical symmetry of the spacetime, the Keplerian frequency $\Omega_{\phi}$ and the vertical epicyclic frequency $\Omega_{\theta}$ coincide and are suppressed by $\alpha$, whereas the radial epicyclic frequency $\Omega_{r}$ is enhanced. The impact of $\alpha$ on several twin-peak HF QPO models is examined, revealing that $\alpha$ increases both the lower and upper bounds of the predicted QPO frequency ranges. By imposing the $2\nu_U = 3\nu_L$ resonance condition, we analyze the resonant radius, upper QPO frequency, maximum allowed black hole mass, and the time delay between the shadow and QPO signals. We find that the resonant radius decreases with $\alpha$, while the upper QPO frequency increases, spanning the range $\nu_U \sim (673/M$–$4360/M)\mathrm{Hz}$. When the Tolman–Oppenheimer–Volkoff limit is imposed, the upper frequency is further constrained to $\nu_U \lesssim 1450$Hz. {Combining astronomical observations for the classification of QPOs, where $\nu_U \geq 100$Hz, the black hole mass in the nonlocal gravity should satisfy $M \lesssim 43.6M_\odot$.} Although the radial separation between the resonant radius and the photon sphere decreases with $\alpha$, the associated gravitational time delay increases, remaining below $\sim 1.3\mathrm{ms}$ and thus negligible for current observational capabilities.
}

\end{abstract}

\maketitle

\section{Introduction}\label{firstpart}
General Relativity (GR) provides a remarkably successful framework for describing the fundamental structure of spacetime and has withstood a wide range of experimental and observational tests. These tests include solar system dynamics \cite{Will:2014kxa,GRAVITY:2020gka}, gravitational wave detection \cite{LIGOScientific:2016aoc,LIGOScientific:2016lio,LIGOScientific:2016sjg,LIGOScientific:2019fpa,LIGOScientific:2025rid}, and imaging of M87* and Sgr A* \cite{EventHorizonTelescope:2019dse,EventHorizonTelescope:2019ths,EventHorizonTelescope:2022xnr,EventHorizonTelescope:2022xqj}. These observations also provide compelling evidence for the existence of black holes.

Despite its empirical success, GR faces several fundamental challenges. The singularity theorems predict that gravitational collapse generically leads to spacetime singularities, characterized by geodesic incompleteness and the breakdown of the classical manifold structure \cite{Penrose:1964wq, Penrose:1969pc, Hawking:1970zqf}. Moreover, gravity is perturbatively non-renormalizable, indicating ultraviolet incompleteness at the quantum level. These features suggest that the classical description of gravity fails in high-energy regimes, motivating the search for a consistent theory of quantum gravity \cite{rovelli2004quantum,Ashtekar:2004eh}.

Several prominent approaches to quantum gravity predict an intrinsic nonlocal structure of spacetime \cite{Masood:2016wma,Modesto:2015ozb,Modesto:2016ofr,Faizal:2017dlb}. In this context, {leading-order quantum corrections can be effectively described as nonlocal modifications of GR \cite{Elizalde:1995tx,Modesto:2010uh,Chicone:2012cc,Joshi:2019cyk},} giving rise to nonlocal gravity (NLG). As a candidate theory of quantum gravity, NLG has attracted considerable attention. It has been shown that appropriately constructed nonlocal theories can simultaneously achieve renormalizability and unitarity, while regularizing curvature singularities in black hole and cosmological spacetimes, thereby addressing the issue of classical spacetime incompleteness \cite{Moffat:2010bh,Biswas:2010zk,Modesto:2011kw,Modesto:2014lga}. By incorporating inverse d’Alembertian operators into the gravitational action \cite{Deser:2007jk,Nojiri:2007uq,Jhingan:2008ym,Capozziello:2008gu,Bahamonde:2017sdo,Deser:2019lmm,Maggiore:2014sia}, NLG models can account for the observed late-time cosmic acceleration without introducing a cosmological constant. Moreover, NLG provides an excellent fit to observational data, yielding predictions consistent with the $\Lambda$CDM model for the cosmic microwave background, baryon acoustic oscillations, and type Ia supernovae \cite{Dirian:2014ara,Dirian:2014bma,Dirian:2016puz}. The spherically symmetric static black hole in NLG has been studied in Refs.~\cite{Nicolini:2012eu,Isi:2013cxa,Frolov:2015bta,Knipfer:2019pgi,DAgostino:2025wgl,Fu:2022yrs}.
 
Astrophysical observations of strong gravitational fields near black holes provide a crucial arena for testing competing theories of gravity. The geodesic motion of test particles encodes essential information about the underlying geometry \cite{Fujita:2009bp,Pugliese:2013zma,hennigar2018,Gao:2023mjb,Shaikh:2019jfr,Zhang:2017nhl,Zhang:2021xhp,Yang:2021chw,Tan:2024hzw,Joshi:2020tlq,Yang:2020jno}. Although the event horizon cannot be observed directly, radiation from the accretion disk carries imprints of the near-horizon structure \cite{Bardeen:1972fi}. In particular, X-ray emission from the inner disk can be used to estimate the location of the innermost stable circular orbit (ISCO).

Ref. \cite{Syunyaev1972} proposed probing strong-field gravity through quasi-periodic oscillations (QPOs) originating from the innermost disk regions. Observationally, QPOs are typically classified into high-frequency (HF, $0.1$–$1$ kHz) and low-frequency (LF, $<0.1$ kHz) components \cite{Stella:1999sj}. HF QPOs often appear as twin peaks (upper and lower modes) with a characteristic frequency ratio close to $3:2$ \cite{Kluzniak:2001ar,Abramowicz:2002xc,Abramowicz:2004rr}. This property has been widely employed to constrain compact-object parameters within various black hole models \cite{Chen:2021jgj,Jiang:2021ajk,Liu:2023vfh,Riaz:2023yde,Rayimbaev:2023bjs,Abdulkhamidov:2024lvp,Jumaniyozov:2024eah,Guo:2025zca,Zhang:2025acq,Sui:2025yem,Shahzadi:2023act}.

Motivated by these developments, we investigate test-particle motion around black holes in NLG, aiming to elucidate the near-horizon spacetime structure through orbital dynamics. We examine circular geodesics and their associated fundamental frequencies, focusing on the influence of the nonlocal parameter on dynamical properties. In particular, we analyze the Keplerian, radial, and vertical epicyclic frequencies and explore their implications for twin-peak HF QPO models. By imposing the resonance condition $2\nu_U = 3\nu_L$, we determine the corresponding orbital radii and derive the range of QPO frequencies and the associated upper bound on the black hole mass consistent with NLG.

The paper is organized as follows. In Section \ref{secondpart}, we briefly review the NLG black hole solution. Section \ref{thirdpart} focuses on the circular motion of test particles around the NLG black hole. In Section \ref{fourthpart}, we calculate the fundamental frequencies of test particle orbits, including the Keplerian frequency and the radial and vertical epicyclic frequencies. Section \ref{qpopart} discusses the characteristic frequencies predicted by various twin-peak HF QPO models. Finally, in Section \ref{conclusion}, we summarize our key findings and present concluding remarks.

\section{A Brief Review of the Black hole in Nonlocal Gravity}~\label{secondpart}
\begin{figure}[htp!]
\includegraphics[width=0.35\textwidth]{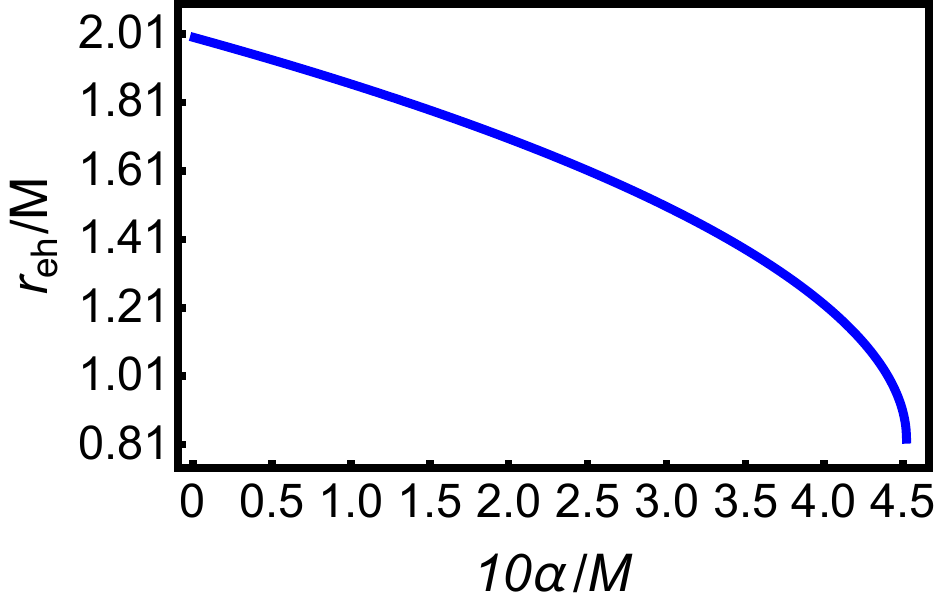}
\caption{The event horizon of the NLG black hole with different nonlocal parameter $\alpha$.}
\label{event horizon}
\end{figure}

In this section, we will give a brief review of the black hole with spherically symmetric form in nonlocal gravity. One can see Refs. \cite{Nicolini:2012eu,Fu:2022yrs} for more details and discussions. For the nonlocal gravity, the action can be set with \cite{Fu:2022yrs,Nicolini:2012eu}:
\begin{eqnarray}
S_{\text{tot}}=\frac{1}{16\pi G}\int d^4x\sqrt{-g}\mathcal{A}^{-2}(\square)R+S_{\text{matt}}, ~\label{action}
\end{eqnarray}
where $\mathcal{A}(\square)$ is a function of dimensionless covariant d'Alembertian operator, i.e., $\square\equiv{\alpha}^{2} g^{\mu\nu}\nabla_{\mu}\nabla_{\nu}$, {and the parameter ${\alpha}$ can be considered as a fundamental length scale of the theory, with the dimension of length}. {\color{blue}This theory is not only ultraviolet complete but also preserves unitarity to all orders in perturbation theory \cite{Modesto:2010uh}.} The corresponding equations of motion can be expressed with
\begin{eqnarray}
\mathcal{A}^{-2}(\square)\left(R_{\mu\nu}-\frac{1}{2}g_{\mu\nu}R\right)=8\pi G T_{\mu\nu}.
\end{eqnarray}
By redefining the effective energy momentum tensor with $\mathcal{T}_{\mu\nu}\equiv \mathcal{A}^{2}(\square)T_{\mu\nu}$, we can rewrite  equations of motion as 
\begin{eqnarray}
R_{\mu\nu}-\frac{1}{2}g_{\mu\nu}R=8\pi G \mathcal{T}_{\mu\nu}.~\label{eom2}
\end{eqnarray}
In this paper, we consider a matter field in the form of a pressureless static fluid, with the $T^0_0$ component of its energy-momentum tensor given by 
\begin{eqnarray}
T^0_0=-\frac{M}{4\pi r^2}\delta(r),
\end{eqnarray}
where $M$ is the mass of the source and $\delta(r)$ is the usual delta function. Then, we assume a static spherically symmetric black hole metric with
\begin{eqnarray}
ds^2=-f(r)dt^2+\frac{1}{f(r)}dr^2+r^2(d\theta^2+\sin^2\theta d\phi^2), ~\label{smetric}
\end{eqnarray}
and the metric component $f(r)$ can be solved as
\begin{eqnarray}
f(r)=1-\frac{2\mathcal{G}(r)M}{r},
\end{eqnarray}
with the effective Newton’s constant expressed  
\begin{eqnarray}
\mathcal{G}(r)=-\frac{4\pi G}{M}\int dr r^2 \mathcal{T}^0_0.
\end{eqnarray}
Here, we should note the explicit expression of $\mathcal{G}(r)$ is contingent upon the selection of $\mathcal{A}(\square)$. In this paper, we will consider the form of $\mathcal{A}(\square)$ with \cite{Nicolini:2012eu}
\begin{eqnarray}
\mathcal{A}(p^2)=\exp (-{\alpha} p/2),
\end{eqnarray}
and the corresponding effective Newton’s constant is
\begin{eqnarray}
\mathcal{G}(r)&=&-\frac{4\pi G}{M}\int dr r^2 \mathcal{T}^0_0\nn\\
&=&\frac{2 G}{\pi}\big(\arctan(r/{\alpha})-\frac{r/{\alpha}}{1+(r/{\alpha})^2}\big).  \label{mofa}
\end{eqnarray}
To ensure that the metric \eqref{smetric} describes a black hole spacetime, it is necessary that $f(r)=0$ has at least one real root. {This necessitates that the nonlocal parameter $\alpha/M\leq0.452$}, and the corresponding event horizon of the NLG black hole is shown in Fig. \ref{event horizon}.

\section{Circular motion of test particle around a NLG black hole }\label{thirdpart}
\subsection{Equation of motion for test particles}
The equation of motion for a test particle with unit mass can be expressed as 
\begin{equation}
\mathcal{L}_{Ld}=\frac{1}{2}g_{\mu\nu}\dot{x}^\mu\dot{x}^\nu, 
\end{equation}
where $\dot{x}^\mu=dx^\mu/\lambda$ is the four-velocity of the test particles and the parameter $\lambda$ is the affine parameter. For this stationary black hole, the Killing vectors $\partial_t$ and $\partial_\phi$ can result the conserved total energy $\mathcal{E}$ and the angular momentum $\mathcal{L}$ as
\begin{equation}
\mathcal{E}=-g_{tt}\dot{t},~~\mathcal{L}=g_{\phi\phi}\dot{\phi}.
\end{equation}
With the normalization condition $g_{\mu\nu}\dot{x}^\mu\dot{x}^\nu=-1$, the four-velocity can be rewritten with 
\begin{eqnarray}
&&g_{rr}\dot{r}^2+g_{\theta\theta}\dot{\theta}^2=-1+\frac{\mathcal{E}^2}{g_{tt}}-\frac{\mathcal{L}^2}{g_{\phi\phi}}=V_{\text{all}},\nn\\
&&V_{\text{all}}=-1+\frac{\mathcal{E}^2}{f(r)}-\frac{\mathcal{L}^2}{r^2\sin(\theta)^2}.\label{eomall}
\end{eqnarray}
or 
\begin{eqnarray}
&&\dot{t}=\frac{\mathcal{E}}{f(r)},~~~~~~~~\dot{\phi}=\frac{\mathcal{L}}{r^2 \sin(\theta)^2},\\ 
&&\dot{r}^2=\mathcal{E}^2-f(r)\Big(1+\frac{\mathcal{K}}{r^2}\Big),\\
&&\dot{\theta}^2=\frac{1}{r^4}\Big(\mathcal{K}-\frac{\mathcal{L}^2}{\sin(\theta)^2} \Big)=V_{\theta},\label{eomtheta}
\end{eqnarray} 
where $\mathcal{K}$ is the Carter constant. 

\begin{figure}[htbp!]
\centering 
\includegraphics[width=0.35\textwidth]{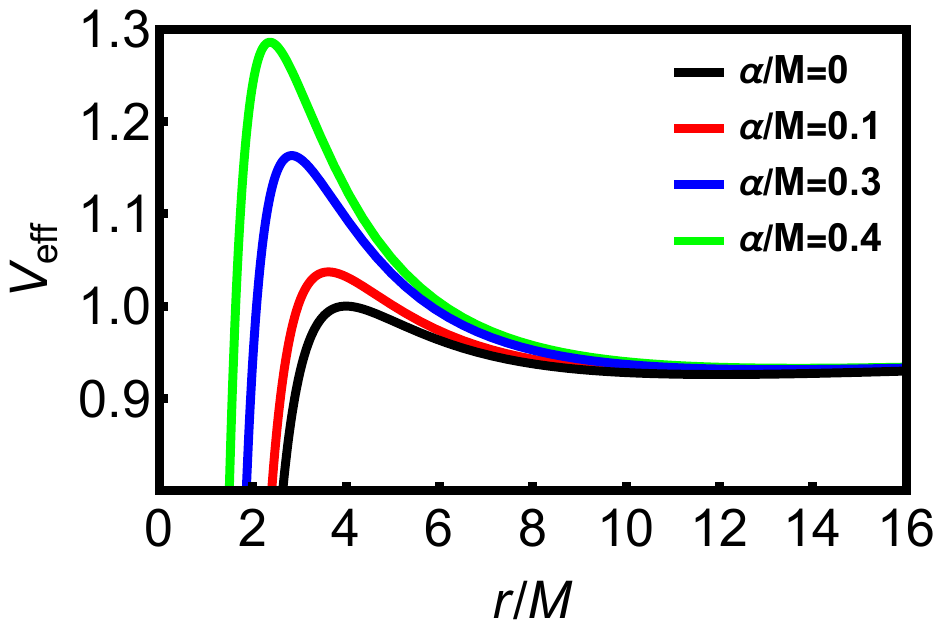}
\caption{Effective potentials $V_{\text{eff}}$ for the radial motion of test particles with angular momentum $\mathcal{L}=4M$, with different values of nonlocal parameter $\alpha$.}\label{effpotential}
\end{figure}

In this paper, we restrict our analysis to equatorial motion, i.e., $\theta=\pi/2$ and $\dot{\theta}=0$, implying $\mathcal{K}=\mathcal{L}^2$. Then, the radial equation of motion can be expresses
\begin{eqnarray} \label{eomr}
&&\dot{r}^2=\mathcal{E}^2-f(r)\Big(1+\frac{\mathcal{L}^2}{r^2}\Big)=\mathcal{E}^2-V_{\text{eff}},\\
&&V_{\text{eff}}=f(r)\Big(1+\frac{\mathcal{L}^2}{r^2}\Big),\nn
\end{eqnarray} 
where $V_{\text{eff}}$ acts as the effective potential for radial motion. Figure~\ref{effpotential} illustrates the influences of nonlocal parameter $\alpha$ on the effective potential $V_{\text{eff}}$. The result shows that the overall height of $V_{\text{eff}}$ increase with $\alpha$. Additionally, the peak of $V_{\text{eff}}$ shifts to smaller radial values as $\alpha$ increases. This behavior indicates that nonlocal parameters exert a significant influence on other kinematic properties of particle motion, which we will explore in the subsequent chapters.

\subsection{Circular orbits}
When a massive test particle moves around the NLG black hole with circular orbits, the corresponding equation of motion in the radial direction should satisfy  
\begin{equation}
\dot{r}=0,~~\ddot{r}=0 ~\Longrightarrow~ V_{\text{eff}}=\mathcal{E}^2,~~\partial_{r}V_{\text{eff}}=0. \label{circular condition}
\end{equation}
With the two condition equations, we can get the expressions of the angular momentum $\mathcal{L}$ and the total energy $\mathcal{E}$ for the circular orbits with
\begin{eqnarray}
&&\mathcal{L}=\sqrt{r^3\partial_r f(r)/\Big(2f(r)-r\partial_rf(r)\Big)},\\
&&\mathcal{E}=\sqrt{2f^2(r)/\Big(2f(r)-r\partial_rf(r)\Big)}.
 \label{circularel}
\end{eqnarray}
The radial dependence of total energy $\mathcal{E}$ and angular momentum $\mathcal{L}$ for the massive  test particles with circular orbits around NLG black hole is shown in Fig. \ref{circularel}. As shown in Fig. \ref{circularel}, both $\mathcal{E}$ and $\mathcal{L}$ decrease with the nonlocal parameter $\alpha$. Furthermore, the presence of the nonlocal parameter $\alpha$ serves to decrease the minima of both energy and angular momentum, concurrently effectuating an inward displacement of the corresponding orbital radii. This critical radius can be interpreted as the radius of the ISCO. Consequently, the ISCO radius diminishes as $\alpha$ escalates.

\begin{figure}[htp!]
\includegraphics[width=0.35\textwidth]{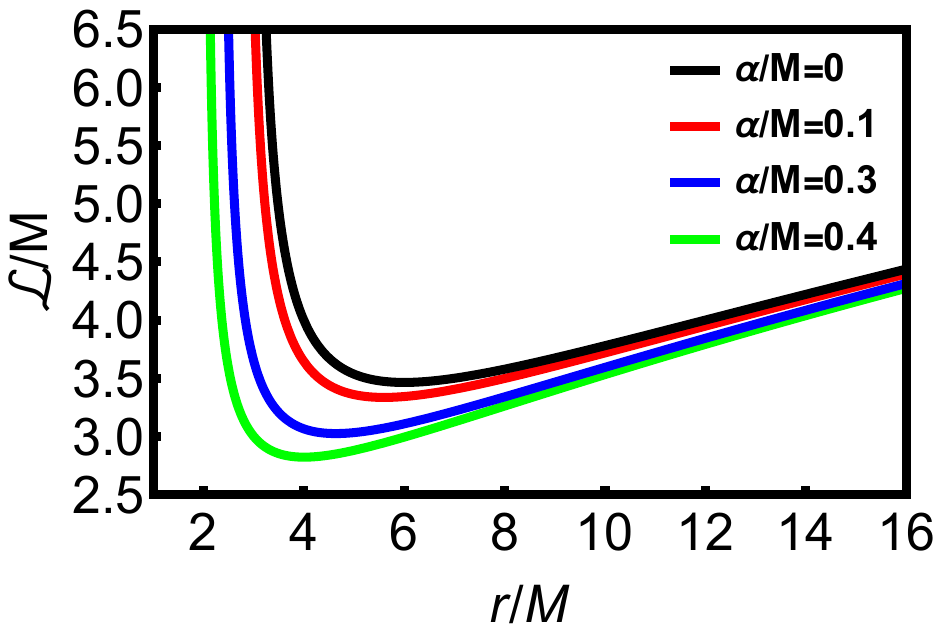}
\includegraphics[width=0.35\textwidth]{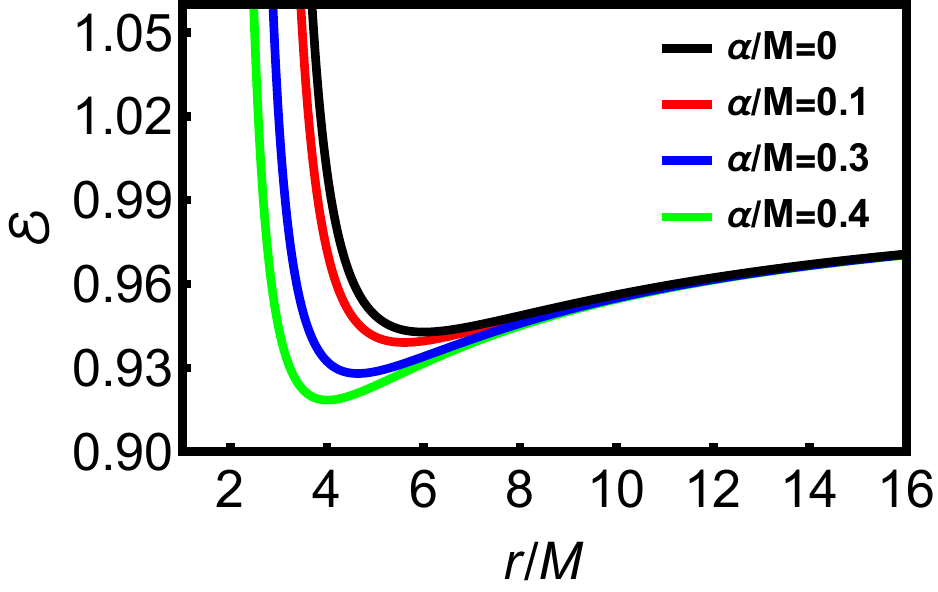}
\caption{The influence of the nonlocal parameter $\alpha$ on the angular momentum $\mathcal{L}$ and specific energy $\mathcal{E}$ of test particles in circular orbits around  the black hole in nonlocal gravity.}
\label{circularel}
\end{figure}

\subsection{Innermost stable circular orbits}
\begin{figure}[htbp!]
{\includegraphics[width=0.35\textwidth]{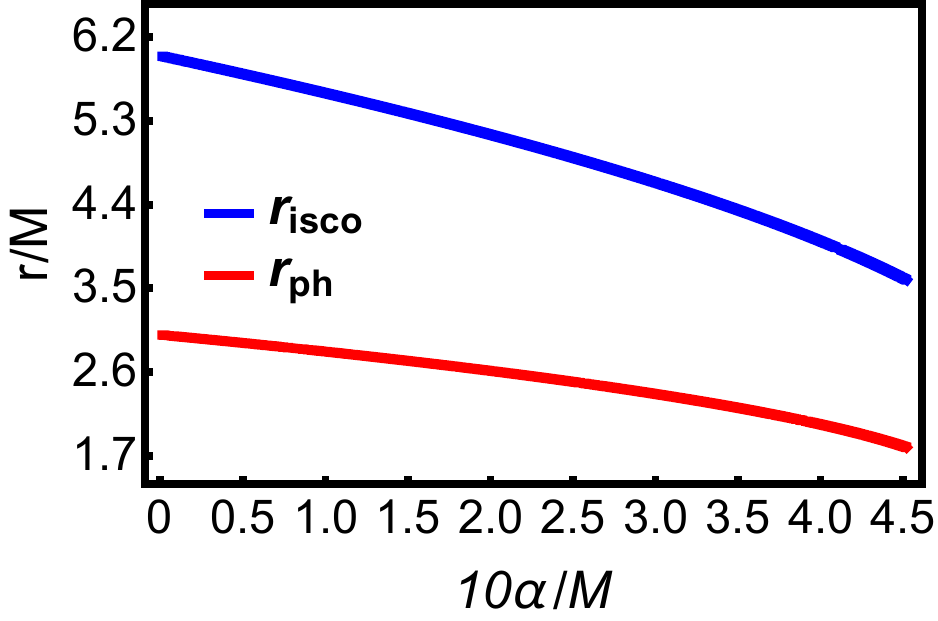}}
\caption{The radii of the ISCO $r_{\text{isco}}$ and photon sphere $r_{\text{ph}}$ for the NLG black hole, with different values of the nonlocal parameter $\alpha$.}\label{iscor}
\end{figure}

\begin{figure}[htp]
\includegraphics[width=0.35\textwidth]{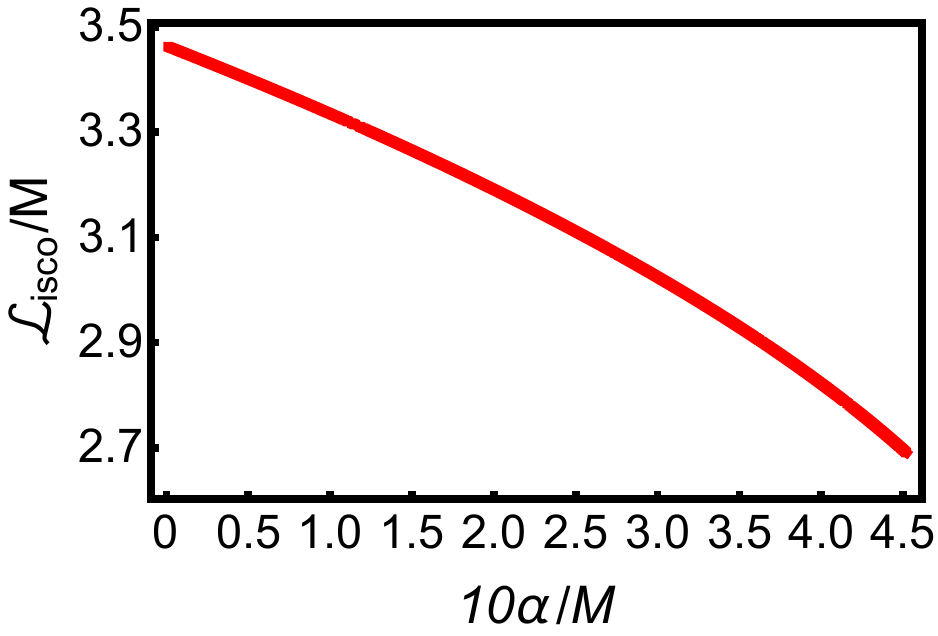}
\includegraphics[width=0.35\textwidth]{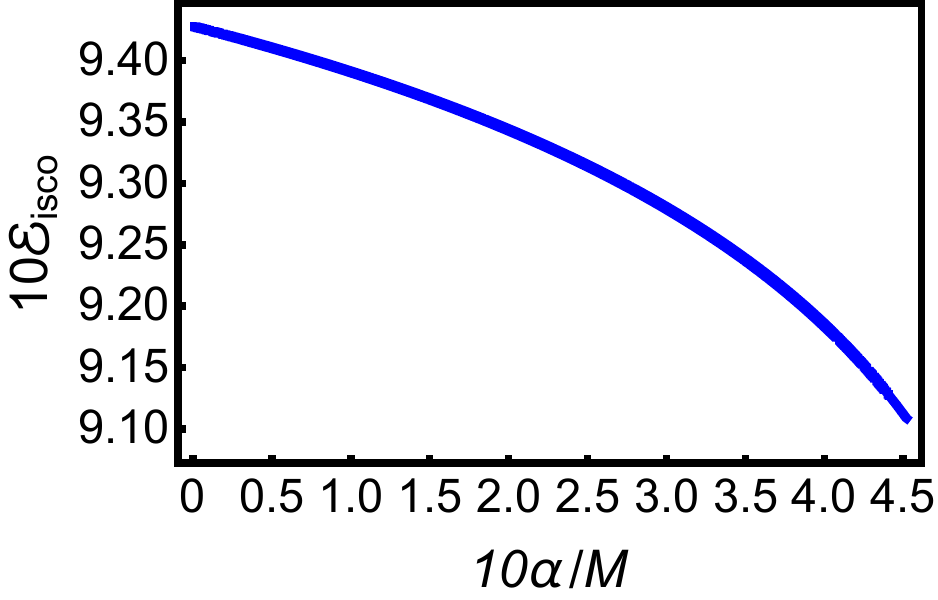}
\caption{The variation of the angular momentum  $\mathcal{L}_{ISCO}$ and total energy $\mathcal{E}_{ISCO}$ at the ISCO with respect to the nonlocal parameter $\alpha$.}
\label{iscole}
\end{figure}

For the stable circular orbits, the second derivative of the effective potential must satisfy the condition: $\partial_r^2 V_{\text{eff}}\geq0$, or equivalently,
\begin{equation}
2r f(r)\partial^2_rf(r)-4 r (\partial_rf(r))^2+6f(r)\partial_rf(r)\geq0.\label{iscocondition}
\end{equation}
Based on the condition $\partial^2_rV_{\text{eff}}=0$, Fig. \ref{iscor} depicts the trends of the ISCO radius $r_{\text{isco}}$ and photon sphere radius $r_{\text{ph}}$ with different $\alpha$. The result shows that both $r_{\text{isco}}$ and $r_{\text{ph}}$ decrease monotonically with $\alpha$. Meanwhile, the interval between $r_{\text{isco}}$ and $r_{\text{ph}}$ decreases concomitantly with $\alpha$. Besides, Fig. \ref{iscole} shows that both the angular momentum $\mathcal{L}_{ISCO}$ and the total energy $\mathcal{E}_{ISCO}$ associated with the ISCO decrease with nonlocal parameter $\alpha$, a trend consistent with that observed in Fig. \ref{circularel}.  

\begin{figure}[htbp!]
{\includegraphics[width=0.35\textwidth]{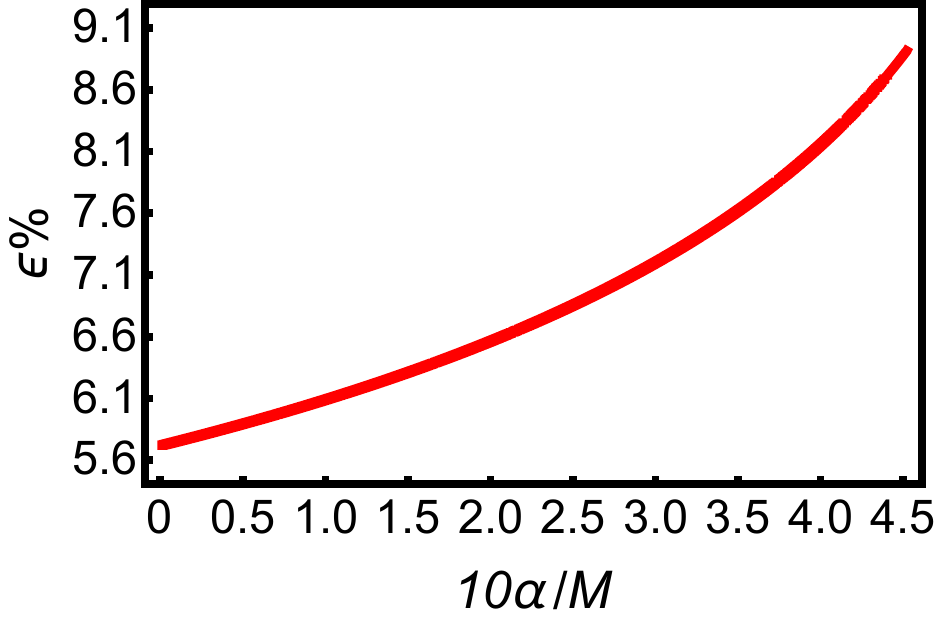}}
\caption{Radiative efficiency $\epsilon $ as a function of the nonlocal parameter $\alpha$.}\label{iscoeff} 
\end{figure}

According to the thin accretion disk model, a test particle falling into a black hole from infinity will radiate energy as it spirals inward \cite{Collodel:2021gxu,Wu:2024sng,Kurmanov:2024hpn,Liu:2024brf}. For a test particle of unit mass, the maximum radiation efficiency $\epsilon$ is defined as the difference between the rest energy and the orbital energy at ISCO. Its expression can be defined with
\begin{equation}
\epsilon=1-\mathcal{E}_{ISCO}, 
\end{equation}
and the corresponding behavior of the radiative efficiency $\epsilon$ varies with the nonlocal parameter $\alpha$ is delineated in Fig. \ref{iscoeff}. As shown, the radiative efficiency $\epsilon$ increases monotonically with $\alpha$, reaching a maximum value of approximately $8.9\%$.

\section{Fundamental frequencies}\label{fourthpart}
In this section, we will study the fundamental frequencies of test particles orbiting a NLG black hole and analyze the influence of the nonlocal parameter $\alpha$ on the Keplerian orbital frequency, as well as the radial and vertical epicyclic frequencies, which characterize oscillations around stable circular orbits.

\subsection{Keplerian frequency}
\begin{figure}[htbp!]
{\includegraphics[width=0.35\textwidth]{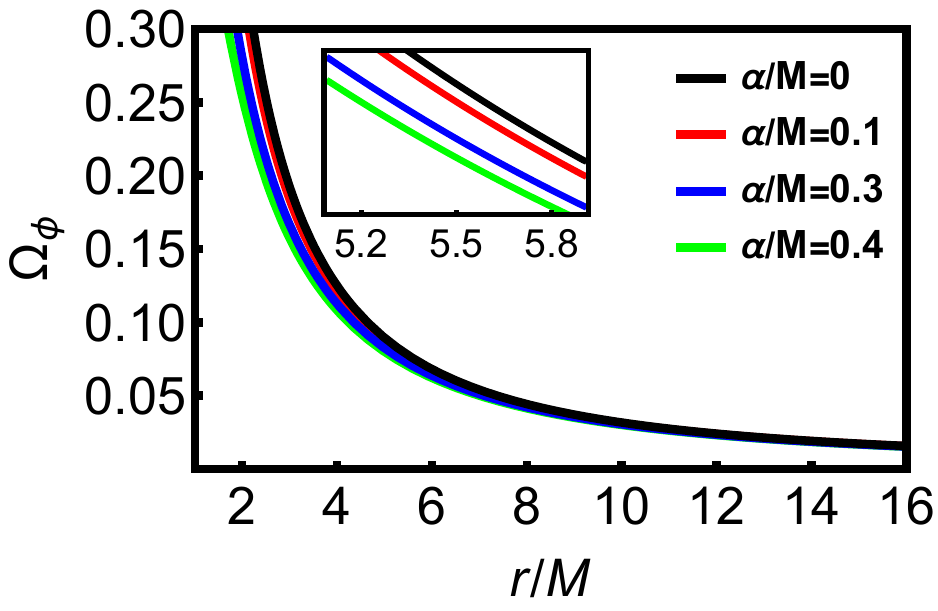}}
\caption{The Keplerian frequency $\Omega_\phi$ of test particles orbiting the NLG black hole is shown as a function of the radial coordinate for different values of the nonlocal parameter $\alpha$.}\label{Kfrequency}
\end{figure}

The angular velocity of a test particle orbiting a black hole, as measured by an observer at infinity, corresponds to the orbital (Keplerian) frequency, $\Omega_\phi = \dot{\phi}/\dot{t}$. In the equatorial plane, $\Omega_\phi$ can be expressed as
\begin{equation}
\Omega_\phi^2=\partial_rf(r)/2r,
\end{equation}
The radial dependence of the Keplerian frequency, $\Omega_\phi$, for test particles with different values of the nonlocal parameter $\alpha$ is shown in Fig.~\ref{Kfrequency}. As illustrated, increasing $\alpha$ accelerates the decrease of $\Omega_\phi$ with the radial coordinate $r/M$, indicating that the nonlocal parameter more strongly suppresses the orbital frequency at larger radii.

\subsection{Harmonic oscillations}
The radial and vertical fundamental frequencies of test particles on circular equatorial orbits can be obtained by introducing small perturbations in the radial and vertical directions, namely $r \to r_0 + \delta r$ and $\theta \to \pi/2 + \delta\theta$, respectively. By imposing the conditions for circular motion, $V_{\text{all}}(r_0,\pi/2) = 0$ and $\partial_{r,\theta} V_{\text{all}} = 0$, the dynamics in the perturbed directions can be well approximated by harmonic oscillations. Consequently, the radial and vertical frequencies, as measured by a distant observer, are determined from the corresponding harmonic oscillator equations:
\begin{eqnarray}
\frac{d^2\delta \theta}{dt^2}+\Omega_\theta^2\delta \theta=0,~~\frac{d^2\delta r}{dt^2}+\Omega_r^2\delta r=0,\label{harmoniceq}
\end{eqnarray}
with 
\begin{eqnarray}
&&\Omega_\theta^2=-\frac{1}{2g_{\phi\phi}\dot{t}^2}\partial^2_{\theta}V_{\text{all}}|_{\theta=\frac{\pi}{2}},\label{hoscillatot}\\
&&\Omega_r^2=-\frac{1}{2g_{rr}\dot{t}^2}\partial^2_{r}V_{\text{all}}|_{\theta=\frac{\pi}{2}},\label{hoscillator}
\end{eqnarray}
the frequencies of the vertical and radial oscillations, respectively. The expressions for the vertical and radial frequencies can be expressed with
\begin{eqnarray}
\Omega_\theta^2&=&\partial_rf(r)/2r=\Omega_\phi^2,\label{oscillatot}\\
\Omega_r^2&=&\frac{1}{2}f(r)\Big(\frac{3\partial_rf(r)}{r}+\partial^2_rf(r)\Big)-\big(\partial_rf(r)\big)^2.\label{oscillator}
\end{eqnarray}

\begin{figure}[htbp!]
{\includegraphics[width=0.35\textwidth]{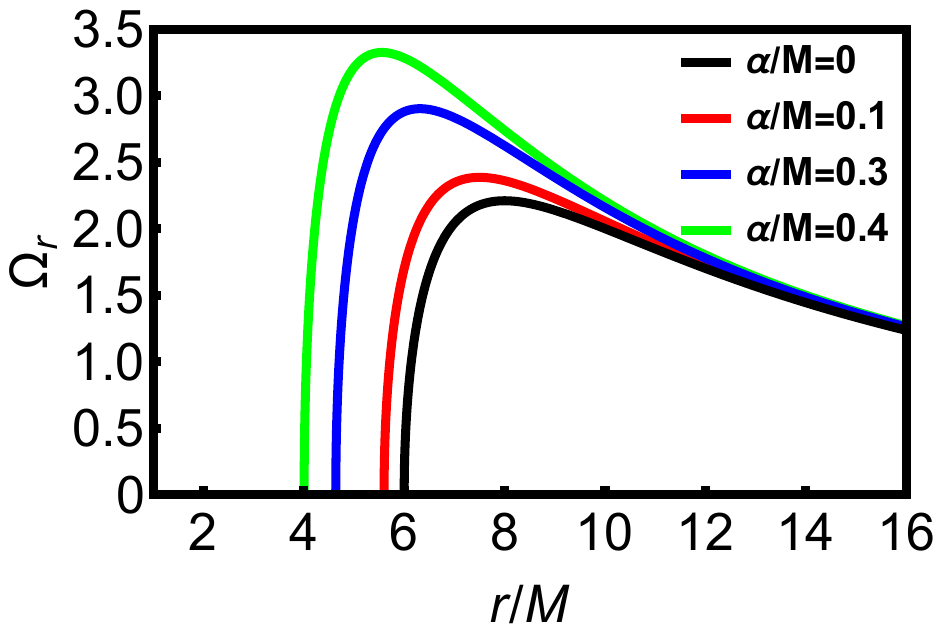}}
\caption{Radial dependence of the radial oscillation frequency $\Omega_r$ of test particles for different values of the nonlocal parameter $\alpha$.}\label{radia oscillation}
\end{figure}

From the results, we see that for this spherically symmetric NLG black hole, $\Omega_{\phi}$ and $\Omega_{\theta}$ share the same expression, meaning that the nonlocal parameter $\alpha$ affects the vertical frequency $\Omega_{\theta}$ in the same way as the Keplerian frequency $\Omega_{\phi}$. Figure~\ref{radia oscillation} illustrates the effect of $\alpha$ on the radial frequency, $\Omega_r^2$. {The results show that the maximum value of the radial frequency $\Omega_r|_{\text{max}}$ increases with $\alpha$, whereas the radial coordinate at which this maximum occurs decreases with $\alpha$.}

\section{Quasi-periodic oscillation around NLG black hole}\label{qpopart}
In this section, we will analyze the predicted frequencies of various twin-peak HF QPO models in the context of an NLG black hole. In most cases, these models emphasize the central role of the black hole’s gravitational field, with the observed QPO frequencies closely related to the fundamental frequencies of harmonic oscillations along geodesic orbits. For static observers at infinity, the fundamental frequencies in physical units (Hz) can be expressed as:
\begin{eqnarray}\label{fundamental frequency}
\nu_{i}=\frac{1}{2\pi}\frac{c^3}{G M}\Omega_{i} [\text{Hz}],~~(i=r,\phi,\theta).
\end{eqnarray}  
We now examine the influence of the nonlocal parameter $\alpha$ on twin-peak HF QPOs around a central black hole, considering various QPO models and different values of the black hole’s physical parameters. In particular, we focus on the following models:
\begin{itemize}
\item The Relativistic Precession (RP) model, originally proposed to explain twin-peak HF QPOs in both black holes and neutron stars \cite{Stella:1999sj}, attributes the observed frequencies to relativistic precession effects. Building on previous work, the RP model can be further subdivided into three additional variants.

\item The Epicyclic Resonance (ER) model explores resonances arising from axisymmetric oscillation modes of thick accretion disks around black holes \cite{Stuchlk:2016,Abramowicz:2001bi}, which are associated with the circular geodesics of test particles. The ER model can also be subdivided into six additional variants.

\item The Warped Disc (WD) model is based on the assumption of non-axisymmetric oscillatory modes within a warped thin accretion disk surrounding black holes \cite{Kato:2004vs,Kato:2008sr}.

\item The Tidal Disruption (TD) model interprets QPOs as manifestations of the tidal disruption of large accreting inhomogeneities \cite{Cadez:2008iv,Kostic:2009hp}.
\end{itemize}
Table \ref{table1} presents the explicit expressions of the upper and lower frequencies of twin-peak HF QPOs associated with the models discussed above. A more detailed exposition of these QPO models can be found in Refs.~\cite{Kotrlova:2014ana,Kolos:2020ykz}. By combining the specific expressions of the various models and noting that $\Omega_{\phi} = \Omega_{\theta}$, we can derive RP0=RP1=RP2=ER1 model and ER3=TD model. This allows the 11 initially distinct models to be reduced to 7 effective models.

\begin{table}[!htb]
\centering
\begin{tabular}{l *{12}{c}}
\hline \hline 
\textbf{model}& \textbf{RP0} & \textbf{RP1} & \textbf{RP2} & \textbf{ER0} & \textbf{ER1} & \textbf{ER2} \\\hline 
$\nu_U$ & $\nu_{\phi}$ & $\nu_{\theta}$ & $\nu_{\phi}$ & $\nu_{\theta}$ & $\nu_{\theta}$ & $\nu_{\theta}-\nu_{r}$ \\\hline 

$\nu_L$ & $\nu_{\phi}-\nu_{r}$ & $\nu_{\phi}-\nu_{r}$ & $\nu_{\theta}-\nu_{r}$ & $\nu_{r}$ & $\nu_{\theta}-\nu_{r}$ & $\nu_{r}$ \\\hline\hline

\textbf{model}& \textbf{ER3} & \textbf{ER4} & \textbf{ER5} & \textbf{WD} & \textbf{TD} \\\hline 
$\nu_U$ &$\nu_{\theta}+\nu_{r}$ & $\nu_{\theta}+\nu_{r}$ & $\nu_{r}$ & $2\nu_{\phi}-\nu_{r}$& $\nu_{\phi}+\nu_{r}$  \\\hline 

$\nu_L$ &$\nu_{\theta}$ & $\nu_{\theta}-\nu_{r}$ & $\nu_{\theta}-\nu_{r}$ & $2\left(\nu_{\phi}-\nu_{r}\right)$ & $\nu_{\phi}$  \\\hline
\end{tabular}
\caption{The explicit expressions of the upper and lower frequencies for the twin-peak HF QPOs associated with RP, ER, WD and TD models.}\label{table1}
\end{table}

\begin{figure}[!htb]
\subfigure[RP and ER5 model]{\label{RPER5 model}
\includegraphics[width=0.23\textwidth]{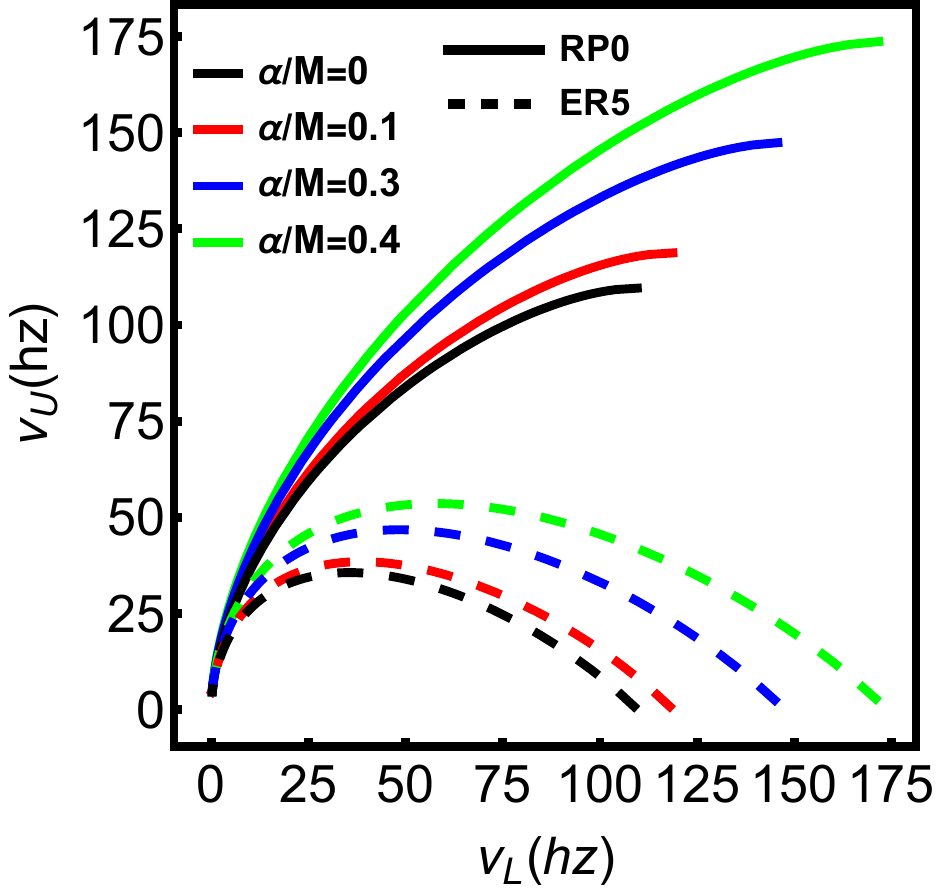}}
\subfigure[ER0 and ER2 model]{\label{ER0ER2 model}
\includegraphics[width=0.23\textwidth]{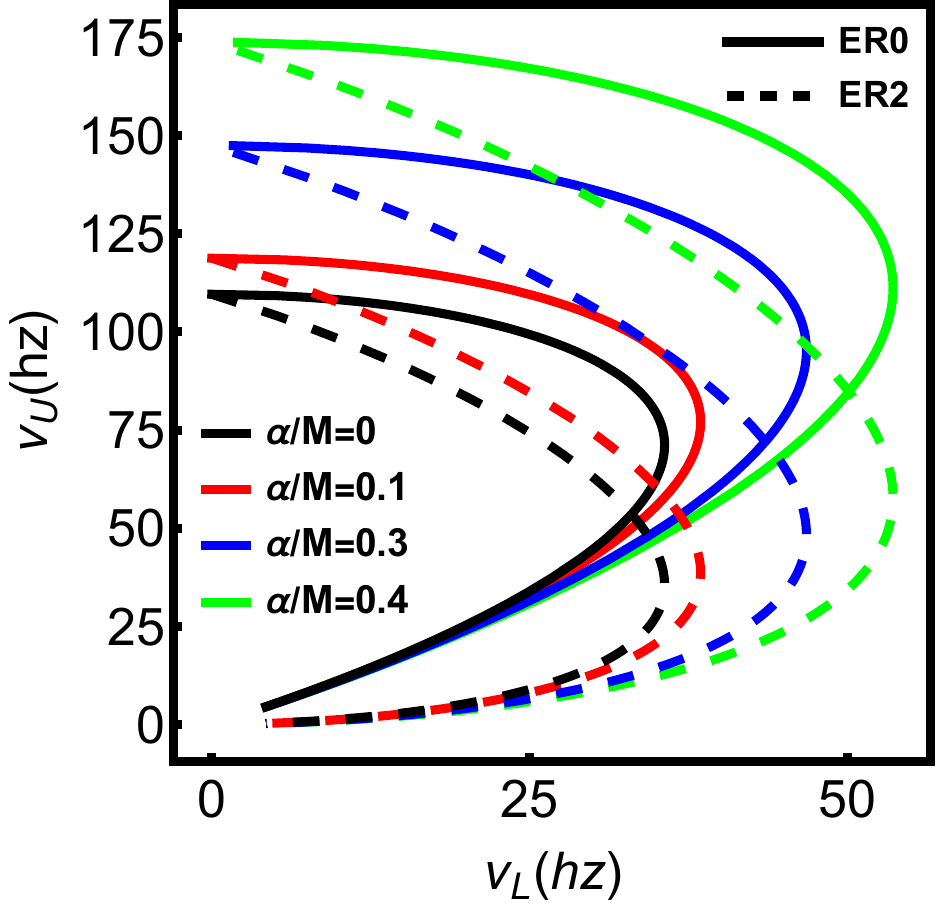}}
\subfigure[ER3 and ER4 model]{\label{ER3ER4 model}
\includegraphics[width=0.23\textwidth]{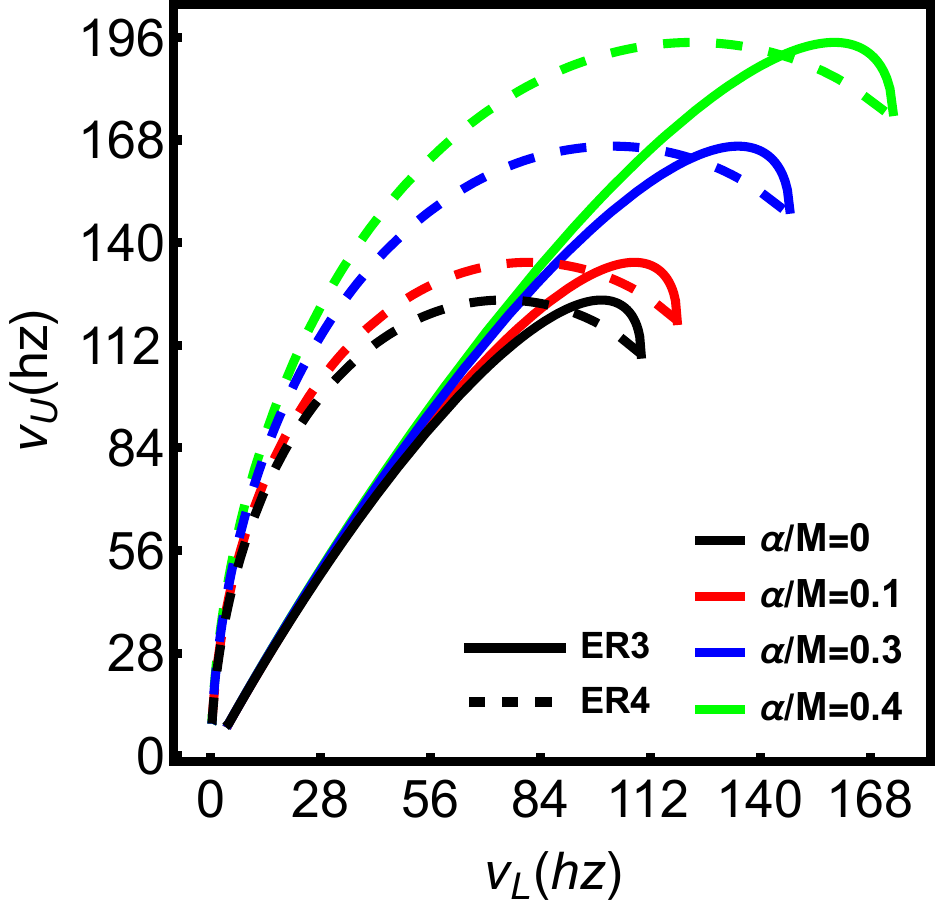}}
\subfigure[WD model]{\label{WD model}
\includegraphics[width=0.23\textwidth]{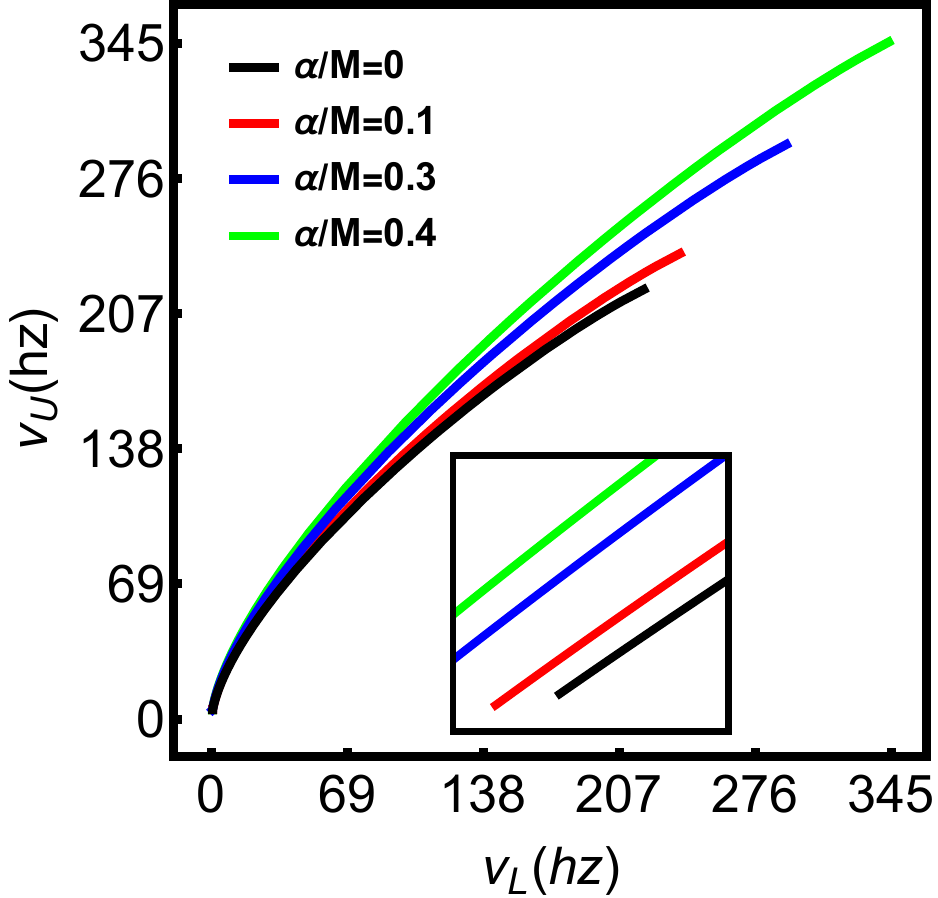}}
\caption{Relations between the upper ($\nu_U$) and lower ($\nu_L$) frequencies of twin-peak HF QPOs in the RP, ER, and WD models around NLG black hole, shown for various values of the nonlocal parameter $\alpha$, assuming the black hole mass with $M=20M_\odot$.}\label{oscillation frequency}
\end{figure}

Figure~\ref{oscillation frequency} shows the relationship between the upper and lower frequencies of twin-peak HF QPOs for the seven models, assuming a black hole mass of $M = 20M_\odot$. From Fig. \ref{oscillation frequency}, we can see that the nonlocal parameter $\alpha$ leads to an expansion of both the upper and lower frequency bounds across all models, indicating that $\alpha$ enhances the QPO frequencies. Notably, the frequencies predicted by the WD model are systematically higher than those of the other models, suggesting that a warped thin disk can amplify QPO frequencies and may provide a plausible explanation for observed high-frequency QPOs.

Furthermore, a model proposed by Abramowicz et al. attributes the origin of twin-peak HF QPOs to resonant phenomena occurring in accretion disks undergoing nearly Keplerian motion. A key implication of this model is that the frequency ratios of these QPOs should be rational. Strong empirical support for this resonance scenario comes from the frequent detection of a $3:2$ frequency ratio ($2\nu_U = 3\nu_L$) in low-mass X-ray binaries  hosting black holes or microquasars \cite{Abramowicz:2002xc,Abramowicz:2004rr}.

\begin{figure}[!htb]
{\includegraphics[width=0.35\textwidth]{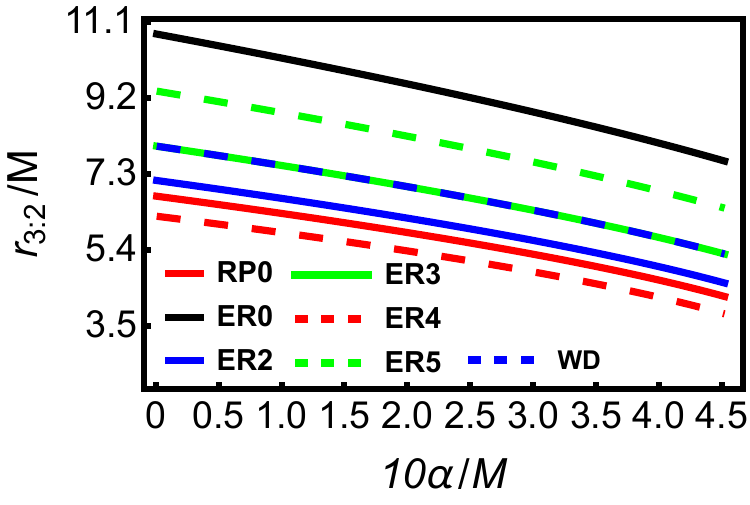}}
\caption{The Resonant radii $r_{3:2}$ corresponding to twin-peak HF QPOs with a $3:2$ frequency ratio, plotted as functions of the nonlocal parameter $\alpha$ for all QPO models.}\label{radii32}
\end{figure}

By imposing the resonance condition $2\nu_U = 3\nu_L$, we can determine the corresponding resonant radius, $r_{3:2}$, associated with twin-peak HF QPOs. Under this condition, the ER3 and WD models yield identical resonant radii. The dependence of the nonlocal parameter $\alpha$ on the resonant radius $r_{3:2}$ is shown in Fig.~\ref{radii32}. As illustrated, $r_{3:2}$ decreases monotonically with $\alpha$ across all QPO models. Among the models considered, the ER0 model predicts resonant radii located farthest from the black hole, whereas the ER4 model consistently yields the smallest radii, closest to the horizon. Despite these quantitative differences, all models exhibit qualitatively similar trends with respect to variations in the nonlocal parameter $\alpha$.

\begin{figure}[!htb]
\includegraphics[width=0.35\textwidth]{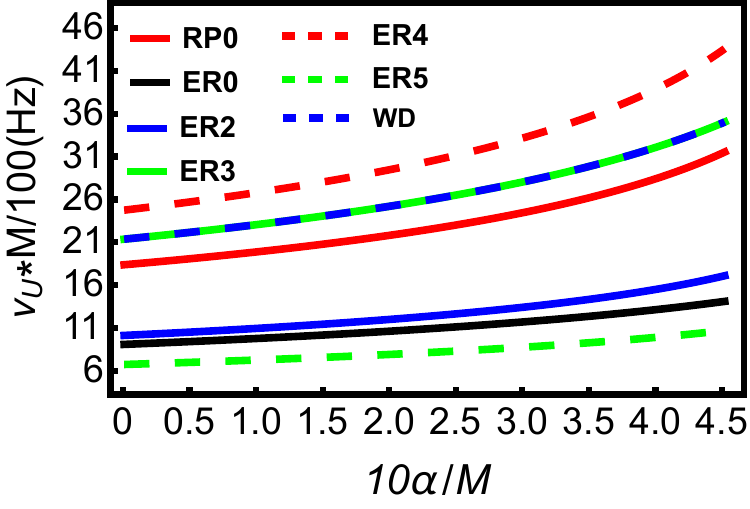}
\caption{Upper frequency $\nu_U$ of twin-peak HF QPOs with the resonant radii $r_{3:2}$ for different QPO models.}\label{upper frequency}
\end{figure}

Figure~\ref{upper frequency} shows the upper HF QPO frequency, $\nu_U$, evaluated at the resonant radius $r_{3:2}$ for different QPO models. As illustrated, $\nu_U$ increases monotonically with the nonlocal parameter $\alpha$ across all models. Under the resonance condition $2\nu_U = 3\nu_L$, the ER3 and WD models yield identical predictions for the upper frequency. Among the models considered, ER4 predicts the highest $\nu_U$ values, whereas ER5 yields the lowest frequencies. Overall, the upper frequency of twin-peak HF QPOs spans approximately $(673/M\text{–}4360/M)$ Hz. By imposing the Tolman–Oppenheimer–Volkoff bound, which sets a lower limit of $3M_{\odot}$ for black hole masses, the corresponding upper limit is $\nu_U \lesssim 1450$ Hz within the NLG black hole framework.

\begin{figure}[htbp!]
\includegraphics[width=0.35\textwidth]{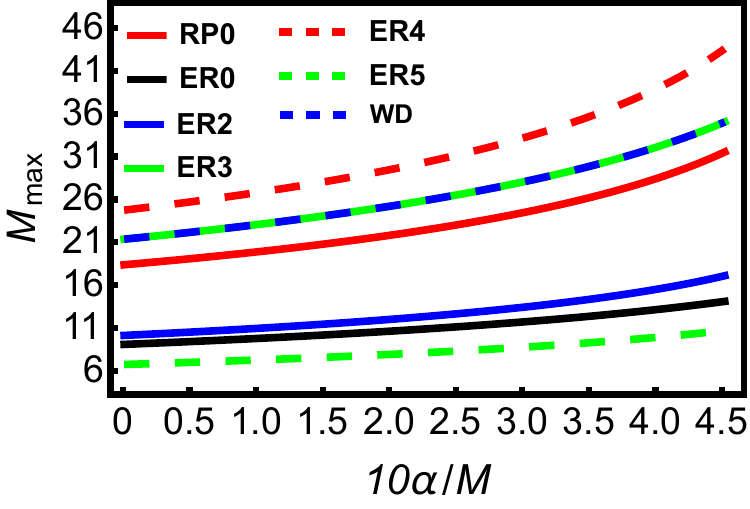}
\caption{The limit of black hole mass $M$ for the twin-peak HF QPOs with the resonant radii $r_{3:2}$ for different QPO models }\label{limit of mass}
\end{figure}

According to the classification of QPOs into LF and HF components, we will adopt $\nu_U = 100$ Hz as an approximate lower bound for HF QPOs, roughly. Consequently, twin-peak HF QPOs are required to satisfy $\nu_U \geq 100$ Hz. By imposing this condition along with the fundamental frequency relation in Eq.~\eqref{fundamental frequency}, we determine the corresponding upper bound on the mass of an NLG black hole, with the results presented in Fig.~\ref{limit of mass}. The results indicate that the maximum allowed mass $M_{max}$ increases monotonically with the nonlocal parameter $\alpha$ across all QPO models. Among the models, ER4 and ER5 exhibit the steepest and mildest growth with respect to $\alpha$, respectively. Overall, within the NLG framework, the black hole mass capable of accounting for twin-peak HF QPOs does not exceed $43.6 M_\odot$.

\begin{figure}[htbp!]
{\includegraphics[width=0.35\textwidth]{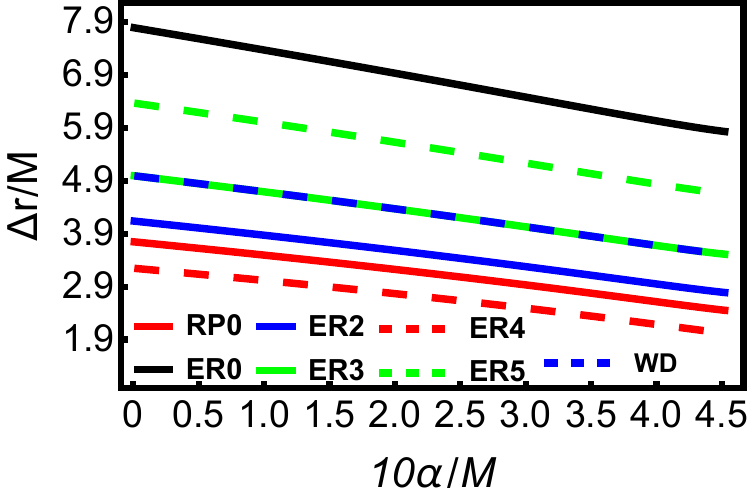}}
\caption{The intervals $\Delta r$ between the resonant radius $r_{3:2}$ and the radius of photon sphere $r_{ph}$ plotted as functions of the nonlocal parameter $\alpha$ for all QPOs models.}\label{drphr32}
\end{figure}

For black holes exhibiting twin-peak HF QPOs, the shadow image provides an additional observable signature. These two signals are associated with the resonant radius $r_{3:2}$ and the photon-sphere radius $r_{ph}$, respectively. The interval $\Delta r$ between the resonant radius $r_{3:2}$ and the photon-sphere radius $r_{ph}$ decreases with $\alpha$ for all QPO models, following an approximately linear trend (see Fig.~\ref{drphr32}).

\begin{figure}[htbp!]{
\includegraphics[width=0.35\textwidth]{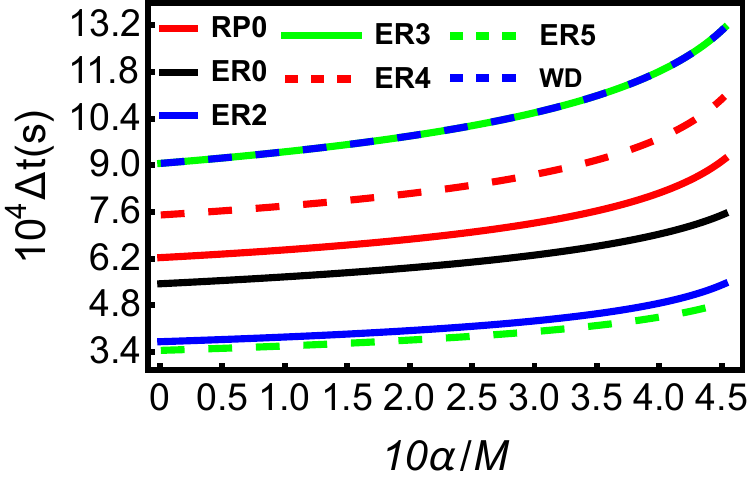}}
\caption{The time delay $\Delta t$ of the QPOs and shadow for the NLG black hole plotted as functions of the nonlocal parameter $\alpha$ for all QPOs models.}\label{timedelay}
\end{figure}

The spatial separation between these characteristic radii also induces a time delay $\Delta t$ between the corresponding signals. For simplicity, we neglect the effects of the accretion disk and cosmological redshift on photon propagation, considering only the gravitational redshift contribution to the time delay, as illustrated in Fig.~\ref{timedelay}. The results show that $\Delta t$ increases with the nonlocal parameter $\alpha$. The ER3 and WD models yield identical predictions and correspond to the largest time delays among all models, whereas the ER2 and ER5 models produce nearly identical results, giving the smallest $\Delta t$. Even in the most extreme case, however, the maximum time delay does not exceed $1.3$ ms, which is negligible for current astronomical observations.

\section{Conclusion and discussion}\label{conclusion}

{In this work, we studied the motion of massive test particles around an NLG black hole and analyzed the associated properties of QPOs, restricting the nonlocal parameter to {$\alpha/M\le 0.452$.}

We find that the nonlocal parameter $\alpha$ can enhance the effective potential $V_{\text{eff}}$, while reduce the specific energy $\mathcal{E}$ and angular momentum $\mathcal{L}$ of circular orbits. The minimum values of $\mathcal{E}$ and $\mathcal{L}$ decrease monotonically with $\alpha$, and the corresponding radii shift inward. Consequently, the ISCO radius, $r_{ISCO}$, decreases with $\alpha$, accompanied by reductions in $\mathcal{E}_{ISCO}$ and $\mathcal{L}_{ISCO}$. In contrast, the radiative efficiency $\epsilon$ increases monotonically, reaching a maximum of approximately $8.9\%$.

We also examined the fundamental orbital frequencies. Owing to the spherical symmetry of the spacetime, the Keplerian frequency $\Omega_\phi$ coincides with the vertical epicyclic frequency, $\Omega_\theta$, and both decrease with $\alpha$. By contrast, the radial epicyclic frequency $\Omega_r$ increases with $\alpha$. For all QPO models, the nonlocal parameter $\alpha$ raises both the lower and upper bounds of the predicted frequency ranges.

Finally, by imposing the resonance condition $2\nu_U = 3\nu_L$, we determined the resonant radius, $r_{3:2}$, the upper frequency, $\nu_U$, the maximum black hole mass, $M_{\rm max}$, and the gravitational time delay, $\Delta t$, between shadow and QPO signals. Under this condition, the ER3 and WD models yield identical predictions. For all models, the resonant radius $r_{3:2}$ decreases with $\alpha$, whereas the corresponding $\nu_U$ increases, spanning approximately $(673/M\text{–}4360/M)$ Hz. Moreover, imposing the TOV limit constrains the upper QPO frequency to $\nu_U \lesssim 1450$ Hz. {Then, based on rough classification of QPOs from astronomical observations, the upper frequency of twin-peak HF QPOs should satisfy $\nu_U\gtrsim100$Hz. Under this condition, the corresponding black hole mass within the NLG framework is constrained to $M \lesssim 43.6 M_\odot$.} Although the radial separation between $r_{3:2}$ and the photon sphere, $r_{ph}$, decreases with $\alpha$, the associated gravitational time delay increases slightly but remains small, $\Delta t \lesssim 1.3$ ms, and is negligible for current observations.
}

\begin{acknowledgments}
{This work is supported with the Scientific Research Foundation for High-level Talents of Anhui University of Science and Technology Grant No. (2024yjrc164)
}.
\end{acknowledgments}


\end{document}